\documentclass[12pt]{article}

\begin{document}

\title{\textbf{ABSENCE OF GLUONIC COMPONENTS IN AXIAL AND TENSOR MESONS}}

\author{\and  W.S. Carvalho\thanks{%
weuber@if.ufrj.br}, A.S. de Castro$^{\S}$\thanks{%
castro@feg.unesp.br}\quad and A.C.B. Antunes\thanks{%
antunes@if.ufrj.br} \\\\
Instituto de F\'\i sica\\
Universidade Federal do Rio de Janeiro\\
C.P. 68528, Ilha do Fund\~ao\\
21945-970 Rio de Janeiro, RJ, Brazil\\
\\
$^{\S }$ UNESP - Campus de Guaratinguet\'{a} - DFQ\\
C.P. 205\\
12500-000 Guaratinguet\'{a}, SP, Brazil}

\date{}
\maketitle

\begin{abstract}
A quarkonium-gluonium mixing scheme previously developed to describe the
characteristic of the pseudoscalar mesons is applied to axial and tensor
mesons. The parameters of the model are determined by fitting the
eigenvalues of a mass matrix. The corresponding eigenvectors give the
proportion of light quarks, strange quarks and glueball in each meson.
However the predictions of the model for branching ratios and
electromagnetic decays are incompatible with the experimental results. These
results suggest the absence of gluonic components in the states of axial and
tensor isosinglet mesons analyzed here.
\end{abstract}

\newpage

\newpage

\section{Introduction}

The existence of gluon self-coupling in QCD gives rise to the possibility of
glueball formation. These states may have the same quantum numbers as those
of some quarkonia. A signature of gluonium states is that they have no place
in the mesons nonets. If a glueball and one or more quarkonia with the same
quantum numbers have nearly the same masses these states may interfere and
new states are formed. Thus the physical states formed by gluonia and
quarkonia interference needs a mixing scheme to describe them. Several kinds
of the mixing schemes has been proposed to give account of the peculiar
properties of these mesons.

In some schemes the physical states are written as linear combinations of
pure quarkonia and gluonia states. The linear coefficients are generally
related to the rotation angles and may be determined by the decay properties
of, or into, the physical mesons \cite{harber}-\cite{closea}.

Another approach, in which the interference is considered at a more
fundamental level, consists in writing a mass matrix for the physical states
in the basis of the pure quarkonia and gluonia states. The elements of this
mass matrix are obtained from a model that describes the process of
interference. The mixtures of the basic states are induced by the
off-diagonal elements. Thus, these elements must contain the amplitudes for
transitions from one to another states of the basis. The eigenvalues of that
matrix give the masses of the physical states and the corresponding
eigenvectors give the proportion of quarkonia and gluonia in each meson \cite
{fritzch1}-\cite{birkel}.

In a previous paper we presented a mixing scheme for the pseudoscalar
mesons, based on a mass matrix approach. The flavor-dependent annihilation
amplitudes and binding energies are the responsible mechanisms for the
quarkonium-gluonium mixing. The properties of the three lowest energy states
of the pseudoscalar isosinglet mesons $\eta (547)$, $\eta ^{\prime }(958)$
and $\eta ^{\prime \prime }(1410)$ are well described by a model based on
the assumption that these states are mixtures of the light quarks, strange
quarks and a glueball \cite{wsc2}.

The nonet of axial ($1^{++},1^3P_1$) and tensor ($2^{++},1^3P_2$) mesons are
well established \cite{PDG}. The axial nonet consists of the isodoublet $%
K_{1A}(1340)$, the isovector $a_1(1260)$ and the isoscalars $f_1(1285)$ and $%
f_1(1510)$. The $K_{1A}$ is a mixture of $K_1(1270)$ and $K_1(1470)$ with a
close to 45$^{\circ }$ mixing angle $\cite{k1}.$ The tensor nonet is formed
by the isodoublet $K_2^{*}(1430)$, the isovector $a_2(1320)$ and the
isoscalars $f_2(1270)$ and $f_2^{\prime} (1525)$. Nonetheless, there are
extra isoscalar states with quantum numbers and masses permitting that they
can be interpreted as partners of the nonets of axial and tensor mesons. The
axial state $f_1(1420)$, observed in two experiments \cite{Gavillet}, has
been considered by some authors \cite{Bityukov} as a possible candidate to
exotic. On the other side, there are two candidate to exotic tensor states: $%
f_2(1640)$ \cite{Alde} and $f_J(1710)$ \cite{Augustin}. There is a
controversy about the value of the spin of the $f_J(1710)$: it may be a
scalar or a tensor state \cite{bes}.

In the present paper the candidates to exotics $f_1(1420)$ and $f_2(1640),$
or $f_2(1710)$, are supposed to be components of quarkonium-gluonium mixing
schemes similar to that previously applied to the pseudoscalar mesons \cite
{wsc2}. The same mixing scheme is not applied to the scalar states because
only the assignment for the scalar isodoublet is well-established.

This paper is organized as follows: The next section outlines a brief review
of the matrix formalism used formerly for the pseudoscalar mesons. We also
fix the notation that will be used in the subsequent sections. The section
three is devoted to the application of the mass matrix formalism for the
three lowest energy states of the axial mesons. Afterwards, in the fourth
section, two different mixing configurations for the tensor isosinglet
mesons are considered. In both sections the results obtained from the mass
matrix formalism are used for calculating some quantities related to
branching ratios and decay widths. Finally, in the conclusion, the results
obtained are analyzed and confronted with those ones presented in the
literature.

\section{The Mass Matrix Formalism}

The mass matrix in the basis $|u\bar u>$, $|d\bar d>$, $|s\bar s>$ and $|gg>$%
, including flavor-dependent binding energies and annihilation amplitudes,
has matrix elements given by 
\begin{equation}
M_{ij}=(2m_i+E_{ij})\delta _{ij}+A_{ij}
\end{equation}
where $i,j=u,d,s,g$. The contribution to the elements of the mass matrix are
: The rest masses of the quarks and the gluon, the eigenvalues $E_{ij}$ of
the Hamiltonian for the stationary bound state ($ij$) and the amplitudes $%
A_{ij}$, that account for the possibility of $q\bar q\leftrightarrow
gg\leftrightarrow q^{\prime }\bar q^{\prime }$ and $q\bar q\leftrightarrow
gg $ transitions. As in the previous paper we assume that $E_{ij}$ and $%
A_{ij}$ are not SU(3)-invariant quantities. Two other bases will be used in
this paper. The first basis consists of the isoscalar singlet and octet of
the SU(3) 
\begin{eqnarray}
|1 > &=& \frac{1}{\sqrt{3}}\left( | u \bar u > + | d \bar d > + | s\bar s >
\right) \\
\nonumber \\
|8 > &=& \frac{1}{\sqrt{6}}\left( | u \bar u > + | d \bar d > - 2 | s\bar s
>\right)
\end{eqnarray}

\noindent The second basis is chosen assuming a segregation of the strange
and the nonstrange quarks 
\begin{eqnarray}
|N > &=& \frac{1}{\sqrt{2}}\left( | u \bar u > + | d \bar d > \right) \\
\nonumber \\
|S > &=& | s \bar s >
\end{eqnarray}

\noindent Besides these states we need also the gluonium and the isovector
states 
\begin{eqnarray}
|G>&=&|gg> \\
\nonumber \\
|\tilde {\pi}^{0}>&=&\frac{1}{\sqrt{2}}(|u \bar u>-|d \bar d>)
\end{eqnarray}

\noindent In these bases the mixing among the isoscalar and isovector states
is caused by isospin symmetry breaking terms. Therefore, assuming the exact
SU(2)-flavor symmetry, one needs only consider the subspace spanned by the
isoscalar states, when the mass matrix reduces to a $3\times 3 $ matrix $M_0$%
.

The invariants of the mass matrix under a unitary transformation give the
following mass relations for the isoscalar physical states: 
\begin{eqnarray}
m_{1}+m_{2}+m_{3} &=&tr(M_{0})  \label{24} \\
&&  \nonumber \\
m_{1}.m_{2}.m_{3} &=&det(M_{0}) \\
&&  \nonumber \\
m_{1}.m_{2}+m_{1}.m_{3}+m_{2}.m_{3} &=&\frac{1}{2}\left[ \left(
tr(M_{0})\right) ^{2}-tr\left( M_{0}^{2}\right) \right]  \label{26}
\end{eqnarray}

\noindent where $M_0$ is the $3\times 3$ mass matrix for the isoscalar
states, and $m_i\;(i=1,2,3)$ are their eigenvalues. \vspace{0.3cm}

The eigenvectors of the mass matrix $M_0$ are the physical states $|m_1>$, $%
|m_2> $ and $|m_3>$ which are mixtures of $|1>$, $|8>$ and $|G>$: 
\begin{eqnarray}  \label{59}
|m_{1}> &=& -\;c_{2}s_{1}\;|{1}>\;+\; c_{1}c_{2}\;|{8}> \;-\;s_{2}\; |G> \\
\nonumber \\
|m_{2}> &=& \;(c_{1}c_{3}+s_{1}s_{2}s_{3})\;|{1}>\;+\;
(c_{3}s_{1}-c_{1}s_{2}s_{3})\;|{8}>\;-\;c_{2}s_{3}\; |G> \\
\nonumber \\
|m_{3}> &=&(c_{1}s_{3}-s_{1}s_{2}c_{3})\;|{1}>\;+\;
(c_{1}s_{2}c_{3}+s_{1}s_{3})\;|{8}> \;+\;c_{2}c_{3}\;|G>  \label{61}
\end{eqnarray}

\vspace{0.3cm} \noindent The coefficients of the eigenvectors are written in
terms of three Euler angles defining a rotation in a three dimensional
space. For brevity, we have defined the notation $c_i\equiv \mathrm{\cos }%
\theta _i$ and $s_i\equiv \mathrm{\sin }\theta _i$ $(i=1,2,3)$.

The eigenvectors (\ref{59})-(\ref{61}) can also be rewritten in the basis $%
|N>$, $|S>$ and $|G>$: 
\begin{eqnarray}
|m_{1}>= &&\;X_{1}|N>\;+\;Y_{1}|S>\;+\;Z_{1}|G>  \label{102} \\
&&  \nonumber \\
|m_{2}>= &&\;X_{2}|N>\;+\;Y_{2}|S>\;+\;Z_{2}|G> \\
&&  \nonumber \\
|m_{3}>= &&\;X_{3}|N>\;+\;Y_{3}|S>\;+\;Z_{3}|G>
\end{eqnarray}

\vspace{0.4cm}

We adopt an expression for the amplitude of the process $q\bar
q\leftrightarrow gg\leftrightarrow q^{\prime }\bar q^{\prime }$ similar to
that of Cohen and Lipkin \cite{lipkina} and Isgur \cite{isgur2}, where the
numerator of the two-gluon annihilation amplitude expression is assumed to
be a SU(3)-invariant parameter, which means that we parameterize the
annihilation amplitude in the form \vspace{0.2cm} 
\begin{eqnarray}
A_{qq^{\prime}}&=&\frac{\Lambda}{m_{q}m_{q^{\prime}}}
\end{eqnarray}

\vspace{0.3cm} \noindent Analogously the amplitude for the processes $q\bar
q\leftrightarrow gg$ is parameterized by 
\begin{eqnarray}
A_{qg}&=&\frac{\Lambda_g}{\sqrt{m_{q}}}
\end{eqnarray}

\noindent according to the results of Close \textit{et al.} \cite{closea}
and K\"{u}hn \textit{et al.} \cite{kuhn}. The phenomenological parameters $%
\Lambda $ and $\Lambda _g$ are to be determined. There is a parameter
relating the binding energies which is very convenient in this mass matrix
formalism, it is defined by 
\begin{equation}
\varepsilon \equiv \frac{1}{2}(E_{uu} + E_{ss}) - E_{us}
\end{equation}
\vskip .5cm \noindent This parameter appears in the formalism when one uses
the basis $|1>$, $|8>$ and $|G>$ (or the basis $|N>$, $|S>$ and $|G>$) and
the mass relation for the non self-conjugate mesons: 
\begin{eqnarray}
m_{I=1/2}&=&m_{u}+m_{s}+E_{us} \\
\nonumber \\
m_{I=1}&=&2m_{u}+E_{uu}
\end{eqnarray}

\noindent The mass matrix contains off-diagonal elements involving not only
the annihilation amplitudes but also the breaking of the SU(3) symmetry in
the binding energies, represented by parameter $\varepsilon$.

The invariants of the mass matrix are functions of $m_s/m_u$, $\Lambda
/m_u^2 $, $\Lambda _g/\sqrt{m_u}$, $\varepsilon$ and $m_G$. These quantities
are not all free. The equations (\ref{24})-(\ref{26}) impose some
constraints among them. These equations can be solved for $m_s/m_u$, $%
\Lambda /m_u^2$ and $\Lambda _g/\sqrt{m_u}$ which are functions of $%
\varepsilon$ and $m_G$. Fixing the values of $\varepsilon$ and $m_G$, the
independent parameters of the model, all the remaining quantities become
determined.

In the pseudoscalar sector \cite{wsc2} the value of $m_{G}$ was limited to
the interval between the masses of the pseudoscalar mesons $\eta $ and $\eta
^{\prime }$, in order to keep the mass matrix Hermitian, because outside
this interval $\Lambda _{g}$ becomes a complex number. For a given value of $%
m_{G}$ the parameter $\varepsilon $ is determined by the minimum of $%
m_{s}/m_{u}$, consistent with the usual values in the nonrelativistic
constituent quark models, which are in the range $1.3-1.8$. For the
determination of $m_{G}$, the remaining free parameter, we searched for the
best values for the data from the branching rations and from electromagnetic
decay widths. We found 
\begin{equation}
m_{s}/m_{u}=1.772
\end{equation}
\noindent and $m_{G}=1300$ MeV. With those values for $m_{s}/m_{u}$ and $%
m_{G}$ we did obtain results for the branching ratios and electromagnetic
decay widths involving the $\eta $, $\eta ^{\prime }$ and $\eta ^{\prime
\prime }$ mesons in reasonable agreement with the experimental data. The
value for the pseudoscalar glueball mass is to be compared with those
predicted by other $\eta -\eta ^{\prime }-\eta ^{\prime \prime }$ mixing
schemes: 1369 MeV \cite{kitamura} and 1302 MeV \cite{genovese}. It must be
observed that the mass of the pseudoscalar glueball given by our model,
similarly to some other mixing schemes is lower than the mass obtained in
the lattice calculations $\sim $ 2300 MeV \cite{teper}-\cite{bali}. In fact
there is an incompatibility between these approaches. Contrarily to what is
obtained in lattice results in the quenched approximation, in the mixing
schemes the pseudoscalar glueball is not assumed to be an isolated physical
state. The mass of the glueball state is obtained simultaneously with the
masses of the $q\bar{q}$ and $s\bar{s}$ pseudoscalar states that are also
components of the physical states. This is probably the source of the
considerable difference between the masses estimated by these approaches.
The ratio $m_{s}/m_{u}$, fixed by the pseudoscalar mesons, will be used as
an input in the axial and tensor sectors.

\section{Axial mesons}

Applying the mixing scheme presented in the previous section to the
isoscalar axial mesons, we find, after fitting the eigenvalues to the
physical masses, the following eigenvectors: 
\begin{eqnarray}
|f_{1}(1285)>&=&0.630\;|1>\;+\;0.735\;|8>\;-\;0.250|G> \\
\nonumber \\
|f_{1}(1420)>&=&-0.391\;|1>\;-\;0.223\;|8>\;-\;0.920|G> \\
\nonumber \\
|f_{1}(1510)>&=&-0.671\;|1>\;+\;0.677\;|8>\;+\;0.302|G>
\end{eqnarray}

\noindent and 
\begin{eqnarray}
|f_{1}(1285)>&=&0.964\;|N>\;+\;0.090\;|S>\;-\;0.250|G> \\
\nonumber \\
|f_{1}(1420)>&=&-0.208\;|N>\;-\;0.332\;|S>\;-\;0.920|G> \\
\nonumber \\
|f_{1}(1510)>&=&0.166\;|N>\;+\;0.939\;|S>\;+\;0.302|G>
\end{eqnarray}

These results suggest that $f_1(1285)$ has $93\%$ of $|N>$, $f_1(1420)$ has $%
85\%$ of $|G>$ and $f_1(1510)$ has $88\%$ of $|S>$. The independent
parameters of the model, corresponding to these eigenvectors, are $%
\varepsilon=25$ MeV and $m_G=1430$ MeV. The remaining parameters are $%
\Lambda /m_u=32.4$ MeV and $\Lambda _g/ \sqrt{m_u}=0.79$ MeV.

The ratio of $J/\psi $ radiative branching ratios into $f_{1}(1420)$ and $%
f_{1}(1285)$ and the ratio of the two-photon width of $f_{1}(1420)$ and $%
f_{1}(1285)$ are given by \cite{seiden}:

\begin{eqnarray}
\frac{B(J/\psi \rightarrow \gamma f_{1}(1420))} {B(J/\psi \rightarrow \gamma
f_{1}(1285))}=\left(\frac{\sqrt{2}X_{2}+Y_{2}} {\sqrt{2}X_{1}+Y_{1}}%
\right)^2 \left(\frac{p_{1}}{p_{2}}\right)^2=\frac{0.85\pm 0.25}{%
B(f_{1}(1420) \rightarrow \eta \pi \pi)} \\
\nonumber \\
\frac{\Gamma_{\gamma \gamma} (f_{1}(1420))} {\Gamma_{\gamma \gamma}
(f_{1}(1285))}=\left(\frac{X_{2}+\frac{\sqrt{2}} {5}Y_{2}}{X_{1}+\frac{\sqrt{%
2}}{5}Y_{1}}\right)^2 \left(\frac{M(f_{1}(1420))}{M(f_{1}(1285))}\right)^2=%
\frac{0.34\pm 0.18} {B(f_{1}(1420)\rightarrow K \bar K \pi)}
\end{eqnarray}

\noindent where $X$ and $Y$ are the mixing coefficients appearing in (14)
and (15) and the labels 1 and 2 stands for the $f_{1}(1285)$ and $%
f_{1}(1420) $, respectively. Our results for those ratios are shown in Table 
\ref{t1} and are to be compared with experimental data.

\section{ Tensor mesons}

The same approach used in the last section is now applied to the tensor
mesons. If we consider the candidate to exotic $f_2(1640)$ as the partner of
the tensor nonet, the resulting mixtures are: 
\begin{eqnarray}
|f_{2}(1270)>&=&0.786\;|1>\;+\;0.480\;|8>\;-\;0.390|G> \\
\nonumber \\
|f_{2}^{\prime}(1525)>&=&0.319\;|1>\;+\;0.598\;|8>\;-\;0.801|G> \\
\nonumber \\
|f_{2}(1640)>&=&-0.642\;|1>\;+\;0.618\;|8>\;+\;0.454|G>
\end{eqnarray}

\noindent On the other hand, we can also consider that it is the $f_2(1710)$
that is mixing with the other tensor isosinglets. In this case we obtain 
\begin{eqnarray}
|f_{2}(1270)>&=&0.360\;|1>\;+\;0.634\;|8>\;-\;0.684|G> \\
\nonumber \\
|f_{2}^{\prime}(1525)>&=&0.746\;|1>\;+\;0.175\;|8>\;+\;0.580|G> \\
\nonumber \\
|f_{2}(1710)>&=&-0.487\;|1>\;+\;0.753\;|8>\;+\;0.442|G>
\end{eqnarray}
Changing to the basis $|N>$, $|S>$, $|G>$ we find that 
\begin{eqnarray}
|f_{2}(1270)>&=&0.919\;|N>\;+\;0.062\;|S>\;-\;0.390|G> \\
\nonumber \\
|f_{2}^{\prime}(1525)>&=&0.371\;|N>\;-\;0.470\;|S>\;-\;0.801|G> \\
\nonumber \\
|f_{2}(1640)>&=&0.134\;|N>\;+\;0.881\;|S>\;+\;0.454|G>
\end{eqnarray}

\noindent and 
\begin{eqnarray}
|f_{2}(1270)>&=&0.726\;|N>\;+\;0.072\;|S>\;-\;0.684|G> \\
\nonumber \\
|f_{2}^{\prime}(1525)>&=&0.602\;|N>\;-\;0.549\;|S>\;+\;0.580|G> \\
\nonumber \\
|f_{2}(1710)>&=&0.336\;|N>\;+\;0.833\;|S>\;+\;0.442|G>
\end{eqnarray}

The independent parameters of the model which give the first set of
eigenstates, $f_2(1270)$, $f_{2}^{\prime}(1525)$ and $f_2(1640)$, are $%
\varepsilon=52$ MeV and $m_G=1510$ MeV. The remaining parameters are $%
\Lambda /m_u^2=-1.0$ MeV and $\Lambda _g/\sqrt{m_u}=4.6$ MeV. The main
content in these states is $85\%$ of $|N>$ in $f_2(1270)$, $64\%$ of $|G>$
in $f_2^{\prime}(1525)$ and $78\%$ of $|S>$ in $f_2(1640)$.

The second set of eigenstates correspond to the following parameters, $%
\varepsilon=-5.0 $ MeV, $m_G=1444$ MeV, $\Lambda /m_u^2=4.8$ MeV and $%
\Lambda _g/\sqrt{m_u}=11 $ MeV. The dominant contribution to each state of
the second set is $53\%$ of $|N>$ in $f_2(1270)$ and $69\%$ of $|S>$ in $%
f_2(1710)$. The states $|N>$, $|S>$ and $|G>$ contribute almost with the
same proportion to the $f_2^{\prime}(1525)$.

The ratio of $J/\psi$ radiative branching ratios into $f_{2}(1270)$ and $%
f_{2}(1525)$ and the ratio of $f(1525)$ branching ratios into $\pi \pi$ and $%
K \bar K$ are given by \cite{seiden} .

\begin{eqnarray}
\frac{B(J/\psi \rightarrow \gamma f_2^{\prime})} {B(J/\psi \rightarrow
\gamma f_2(1270))}=\left(\frac{\sqrt{2}X_{2}+Y_{2}} {\sqrt{2}X_{1}+Y_{1}}%
\right)^2 \left(\frac{p_{2}}{p_{1}}\right)^3 \\
\nonumber \\
\frac{B(f_{2}^{\prime}\rightarrow \pi \pi )} {B(f_{2}^{\prime}\rightarrow K
\bar K)}=\frac{3X_{2}^2} {2(\frac{X_{2}}{\sqrt{2}}+Y_{2})^2} \left(\frac{%
p_{\pi}}{p_{K}}\right)^5
\end{eqnarray}

\noindent Where $X$ and $Y$ are the mixing coefficients appearing in (14)
and (15) and the labels 1 and 2 refers to $f_2(1270)$ and $%
f_{2}^{\prime}(1525)$, respectively. Our results for those ratios are shown
in Table \ref{t1} and are to be compared with experimental data.

\section{Conclusion}

M. Birkel and H. Fritzsch \cite{birkel} have used SU(3)-invariant
annihilation amplitudes in a quadratic mass matrix formalism for describing
the mixing in the axial sector. They found that the candidate to exotic $%
f_1(1420)$ has a gluonic content of 58\% and the gluonic component with a
mass of 1432 MeV. These results are to be compared to the ones found in our
mixing schemes. We found a gluonic content of 85\% in the $f_1(1420)$ and a
gluonic component at 1430 MeV.

In the tensor sector we found that the $f_{2}(1270)$ could be mainly an $|N>$
state. Nevertheless, we found that the candidate to exotic $f_{2}(1640)$ is
predominantly a $|S>$ state, whereas the $f_{2}^{\prime }(1525)$ is mainly a 
$|G>$ state. These results are in contrast with those found in the
literature that indicate the $f_{2}^{\prime }(1525)$ and $f_{2}(1640)$ are
mainly $|S>$ and $|G>$ states, respectively \cite{caruso}, \cite{rosner}-%
\cite{teshima} . Our results were obtained for a gluonic component at 1510
MeV. On the other side, if the physical state is the $f_{2}(1710)$ we found
that it is mainly a $|S>$ state and $f_{2}^{\prime }(1525)$ is nearly an
equiprobable distribution among $|N>$, $|S>$ and $|G>$. For this set of
eigenstates a gluonic component at 1444 MeV was obtained. In the first set
of eigenstates, the mass of the gluonic component is comparable to the mass
found in the range 1536-1590 MeV obtained by other mixing schemes \cite
{schnitzer}.

Here, as in the case of the pseudoscalar sector, we have obtained masses for
a glueball state lower than the obtained in the quenched lattice ($\sim $
2000-2300 MeV) \cite{bali}-\cite{chen}. The source of the substantial
difference among the masses is probably the same as that in the case of the
pseudoscalar mesons: The mass of the glueball states in the mass matrix
formalism is obtained regarding the glueball as being a component of a
physical state, whereas in the lattice calculations the glueball is a
physical state itself. Nevertheless, the results given by the present
quarkonium-gluonium mixing scheme for branching ratios and electromagnetic
decay widths involving the axial and tensor mesons $f_{1}(1285)$, $%
f_{1}(1420)$, $f_{2}(1270)$ and $f_{2}(1525)$ are in clear contradiction
with the experimental ones. The theoretical and experimental results are
compared in Table \ref{t1}. These results show that the quarkonium-gluonium
mixing model, which works well for scalar isosinglet mesons \cite{wsc2}, is
not compatible with the constraints coming from decays concerning the axial
and tensor isosinglet mesons considered in this work. 

The incompatibility above mentioned point out that the presence of gluonic
components in the axial and tensor isosinglet meson states considered here
may be a wrong assumption. On the other hand the interpretation of the states%
\textbf{\ $f_{1}(1420)$, $f_{1}(1510)$, $f_{2}(1640)$ }and\textbf{\ $%
f_{J}(1710)$ }are  controversial  and moreover some of them needs
confirmation \cite{PDG}{.}

\begin{table}[tbp]
\caption{Branching ratios and electromagnetic decay widths involving the $%
f_1 (1285)$, $f_1 (1420)$, $f_2 (1270)$ and $f_2 (1525)$. The results for
tensor mesons were obtained in view of $f_{2}(1640)$ or $f_{2}(1710)$ as
member of the tensor nonet: The results in parenthesis refers to $%
f_{2}(1710) $. Our results are compared with the experimental data.}
\label{t1}
\begin{center}
\begin{tabular}{ccc}
\hline\hline
Observable & Our Model & Experiment \cite{PDG} \\ \hline
&  &  \\ 
$\frac{B(J/\psi \rightarrow \gamma f_{1}(1420))} {B(J/\psi \rightarrow
\gamma f_{1}(1285))}$ & 0.16 & 1.7--8.5 \\ 
&  &  \\ 
$\frac{\Gamma_{\gamma \gamma} (f_{1}(1420))} {\Gamma_{\gamma \gamma}
(f_{1}(1285))}$ & 0.13 & 0.34--0.68 \\ 
&  &  \\ 
$\frac{B(J/\psi \rightarrow \gamma f_2^{\prime})} {B(J/\psi \rightarrow
\gamma f_2(1270))}$ & 0.0012 (0.06) & 0.19 \\ 
&  &  \\ 
$\frac{B(f_{2}^{\prime}\rightarrow \pi \pi )} {B(f_{2}^{\prime}\rightarrow K
\bar K)}$ & 17 (127) & 0.096 \\ 
&  &  \\ \hline\hline
\end{tabular}
\end{center}
\end{table}

\vspace{1.0cm}

\noindent \textbf{Acknowledgments:} This work was partly supported by CNPq,
FINEP and FAPESP. The authors are grateful to the anonymous referee for his
valuable criticism. \vfill\eject

\end{document}